\newif\ifblind
\newif\ifmscclpp
\newif\ifarxiv

\blindfalse
\mscclpptrue
\arxivfalse

\documentclass[sigplan,screen]{acmart}

\usepackage{algorithm}
\usepackage{amsmath}
\usepackage{bm}
\usepackage{subcaption}
\usepackage{svg}
\usepackage{makecell}
\usepackage[toc]{appendix}
\usepackage{listings}
\usepackage{xcolor}
\usepackage{soul}
\usepackage{rotating}
\usepackage{pifont}
\usepackage{multirow}
\usepackage{enumitem}
\usepackage{framed}
\usepackage{tikz}
\usepackage[capitalise,nameinlink]{cleveref}
\usepackage{seqsplit}
\usepackage{fancyvrb}

\ifmscclpp
\newcommand{\mscclpp}[1]{\mbox{MSCCL++}}
\newcommand{\specl}[1]{\mbox{MSCCL++}}
\else
\newcommand{\mscclpp}[1]{\mbox{SpeCL}}
\newcommand{\specl}[1]{\mbox{SpeCL}}
\fi
\ifblind
\newcommand{\aj}[1]{#1} 
\newcommand{\zy}[1]{#1} 
\else
\newcommand{\aj}[1]{}
\newcommand{\zy}[1]{}
\fi
\newcommand{\spara}[1]{\noindent\textbf{#1}}

\newcommand{\blue}[1]{#1}

\newcommand{\footnotetrim}[1]{}

\newcommand{\trimNew}[1]{}
\newcommand{\extended}[1]{}

\RequirePackage[squaren]{SIunits}
\newcommand{\usec}[1]{\SImu s}

\newcommand*\circled[1]{\tikz[baseline=(char.base)]{
            \node[shape=circle,draw,inner sep=0.8pt] (char) {#1};}}
\definecolor{commentgreen}{RGB}{2,112,10}

\copyrightyear{2026}
\acmYear{2026}
\setcopyright{cc}
\setcctype{by}
\acmConference[ASPLOS '26]{Proceedings of the 31st ACM International Conference on Architectural Support for Programming Languages and Operating Systems, Volume 2}{March 22--26, 2026}{Pittsburgh, PA, USA}
\acmBooktitle{Proceedings of the 31st ACM International Conference on Architectural Support for Programming Languages and Operating Systems, Volume 2 (ASPLOS '26), March 22--26, 2026, Pittsburgh, PA, USA}
\acmPrice{}
\acmDOI{10.1145/3779212.3790188}
\acmISBN{979-8-4007-2359-9/2026/03}


\begin{document}

\date{}

\title{\mscclpp{}: Rethinking GPU Communication Abstractions for AI Inference}

\author{Changho Hwang}
\affiliation{%
  \institution{Microsoft Research}
  \city{Vancouver, BC}
  \country{Canada}
}

\author{Peng Cheng}
\affiliation{%
  \institution{Microsoft Research}
  \city{Redmond, WA}
  \country{USA}
}

\author{Roshan Dathathri}
\affiliation{%
  \institution{Microsoft Research}
  \city{Redmond, WA}
  \country{USA}
}

\author{Abhinav Jangda}
\affiliation{%
  \institution{Microsoft Research}
  \city{Redmond, WA}
  \country{USA}
}

\author{Saeed Maleki}
\authornote{Now at xAI.}
\affiliation{%
  \institution{Microsoft Research}
  \city{Redmond, WA}
  \country{USA}
}

\author{Madan Musuvathi}
\affiliation{%
  \institution{Microsoft Research}
  \city{Redmond, WA}
  \country{USA}
}

\author{Olli Saarikivi}
\authornote{Now at Apple.}
\affiliation{%
  \institution{Microsoft Research}
  \city{Redmond, WA}
  \country{USA}
}

\author{Aashaka Shah}
\affiliation{%
  \institution{Microsoft Research}
  \city{Redmond, WA}
  \country{USA}
}

\author{Ziyue Yang}
\affiliation{%
  \institution{Microsoft Research}
  \city{Beijing}
  \country{China}
}

\author{Binyang Li}
\affiliation{%
  \institution{Microsoft Azure}
  \city{Redmond, WA}
  \country{USA}
}

\author{Caio Rocha}
\affiliation{%
  \institution{Microsoft Azure}
  \city{Redmond, WA}
  \country{USA}
}

\author{Qinghua Zhou}
\affiliation{%
  \institution{Microsoft Azure}
  \city{Redmond, WA}
  \country{USA}
}

\author{Mahdieh Ghazimirsaeed}
\affiliation{%
  \institution{Microsoft Azure}
  \city{Cambridge, MA}
  \country{USA}
}

\author{Sreevatsa Anantharamu}
\affiliation{%
  \institution{Microsoft Azure}
  \city{Minneapolis, MN}
  \country{USA}
}

\author{Jithin Jose}
\affiliation{%
  \institution{Microsoft Azure}
  \city{Austin, TX}
  \country{USA}
}

\renewcommand{\shortauthors}{Changho Hwang et al.}

\begin{abstract}
AI applications increasingly run on fast-evolving, heterogeneous hardware to maximize performance, but general-purpose libraries lag in supporting these features. 
Performance-minded programmers often build custom communication stacks that are fast but error-prone and non-portable. 

This paper introduces \textbf{MSCCL++}, a design methodology for developing high-performance, portable communication kernels. 
It provides (1) a low-level, performance-preserving primitive interface that exposes minimal hardware abstractions while hiding the complexities of synchronization and consistency, 
(2) a higher-level DSL for application developers to implement workload-specific communication algorithms, and 
(3) a library of efficient algorithms implementing the standard collective API, enabling adoption by users with minimal expertise.  

Compared to state-of-the-art baselines, MSCCL++ achieves geomean speedups of $1.7\times$ (up to $5.4\times$) for collective communication and $1.2\times$ (up to $1.38\times$) for AI inference workloads.
MSCCL++ is in production of multiple AI services provided by Microsoft Azure, and has also been adopted by RCCL, the GPU collective communication library maintained by AMD. MSCCL++ is open source and available at \url{https://github.com/microsoft/mscclpp}.
Our two years of experience with MSCCL++ suggests that its abstractions are robust, enabling support for new hardware features, such as \texttt{multimem}, within weeks of development.
\end{abstract}

\maketitle

\section{Introduction}

GPU communication has emerged as a critical bottleneck and an opportunity for optimization in \trimNew{high-performance} AI systems. 
For instance, communication kernels account for approximately 10\%-40\% in many real-world Large Language Model (LLM) inference tasks~\cite{nvidia2025jax, gond2025tokenweaveefficientcomputecommunicationoverlap, tutel}.
Recent research has addressed communication bottlenecks, including developing efficient communication algorithms (e.g., routing paths and transfer scheduling)~\cite{taccl,crux,te-ccl}, overlapping communication with computation~\cite{coconet,WangWSDIHCMMZKG23}, and other lower-level optimizations in the stack~\cite{nccl,rccl}. 

Despite these efforts, achieving high communication performance is still challenging and time-consuming in real applications.
A common practice today is for practitioners to write custom communication code from scratch to achieve maximum performance without using standard libraries such as the NVIDIA Collective Communication Library (NCCL)~\cite{nccl}. 
For instance, NVIDIA's TensorRT-LLM~\cite{trt-llm} implements custom AllReduce communication kernels rather than relying on or updating NCCL. There are multiple reasons for doing so.  
First, achieving peak performance requires a careful balance between latency and bandwidth, tailored to the specific message sizes and communication patterns of a workload --- something a general-purpose library, designed to support many workloads, cannot always optimize~\cite{cowan2023mscclang}. For example, TensorRT-LLM uses a custom AllReduce that outperforms NCCL on small data sizes, while falling back to NCCL for larger ones. Second, communication collectives often involve computations (e.g., reductions), and their performance depends on exploiting parallelism in ways that interact well with the necessary computation, further motivating workload-specific customization. Third, the enormous computational demand of modern AI applications is driving rapid evolution in both chips and interconnects, and practitioners rush to exploit these new capabilities long before general-purpose libraries such as NCCL are fully adapted.

The practice of writing custom communication code comes at a huge engineering cost. Developing these libraries is challenging and error prone. New hardware capabilities are often immature or poorly documented, forcing developers to navigate low-level details with limited guidance. There are also complex synchronization and data consistency challenges that arise from interactions across GPUs, CPUs, and NICs, which must be carefully managed to avoid correctness issues and performance bottlenecks. Finally, achieving maximum efficiency typically demands the design of custom communication algorithms tailored to the specific workload and hardware, further increasing development complexity.

\blue{Our hypothesis to address these issues is that \textit{multi-layered} programming abstractions are needed to provide performance, portability, and productivity at the same time.}
This paper proposes
\textbf{\mscclpp{}}\footnote{\mscclpp{} stands for
\underline{M}icro\underline{s}oft \underline{C}ollective \underline{C}ommunication \underline{L}ibrary \underline{++}, pronounced \textit{em-sickle plus-plus}.},
\blue{a design methodology that evaluates this hypothesis for high-performance GPU communication.}
\mscclpp{} provides portable, hierarchical abstractions that enable users to explore the trade-offs between productivity and performance.
At its core, \mscclpp{} provides a Primitive API that exposes minimal performance-preserving abstractions of hardware interconnect methods. These abstractions
provide full flexibility for customization, including fine-grained overlapping of compute and communication, while hiding the complexities of synchronization and ensuring portability across hardware vendors and generations.
Building on the Primitive API, \mscclpp{} provides a high-level Domain-Specific Language~(DSL) 
for specifying custom communication algorithms while the \mscclpp{} DSL Executor handles implementation concerns and performs optimizations when executing the DSL program using the Primitive API. 
Using the DSL, \mscclpp{} provides a library of collective algorithms optimized for different scenarios that together provide the same API as NCCL.
Our design is thus based on a separation of concerns:
(1) the Primitive API provides a portable substrate
for expert programmers to exploit hardware-specific features
for the best performance;
(2) the DSL API offers a streamlined interface that allows application developers to implement workload-specific
and network-topology-specific optimizations with minimal effort;
and (3) the Collective API enables users to leverage
these specialized implementations with no additional effort.
\blue{These three layers offer abstractions that improve programmability while enabling progressive customization layer-by-layer to get the last ounce of performance.}

Prior primitive interfaces not only are hardware-vendor-specific but also often mask certain hardware features, limiting opportunities for optimization; e.g., NCCL provides a low-level {\it send-recv} abstraction that prohibits developers from exploiting one-sided, asynchronous communication over interconnects, and NVSHMEM~\cite{nvshmem} provides a low-level shared-memory abstraction that prevents interconnect-specific optimizations such as using port-mapped DMA-copy on NVLink. 
In contrast, the \mscclpp{} Primitive API exposes the fundamental building blocks of communication for different data transfer modes supported by hardware interconnects: memory-mapped I/O, port-mapped I/O, and switch-mapped I/O. 
While the building blocks are specific to the data transfer modes, they are portable across hardware vendors and generations supporting the corresponding data transfer mode.
This interface supports one-sided and asynchronous abstractions for efficient communication both within and across nodes. The implementation of these primitives handles orchestration and synchronization across GPU, CPU, and NIC, providing clear data consistency and synchronization semantics. 

The \mscclpp{} DSL allows developers to specify communication algorithms while abstracting away implementation details.
To improve programmability, it provides a thread-block-based API with a global view across all GPUs and exposes the underlying 
one-sided, asynchronous communication abstractions of the Primitive API.
\mscclpp{} automatically inserts intra-thread-block synchronizations, fuses instructions, and removes unnecessary memory accesses before lowering to 
an execution plan that is interpreted by the \mscclpp{} DSL Executor. 
This design accelerates the development of application-specific communication algorithms for new hardware environments.
Unlike MSCCLang~\cite{cowan2023mscclang}, our DSL retains the one-sided and asynchronous properties of the Primitive API and enables algorithms that overlap communication with computation.
Using the DSL, we implement efficient communication algorithms, 
including those with computation-communication overlap,
that are tuned for AI inference workloads. 
These algorithms are bundled into our library that provides the 
\mscclpp{} Collective API.

In our evaluation, we compare \mscclpp{} with NCCL,
ROCm Collective Communication Library (RCCL)~\cite{rccl} and Microsoft Collective Communication Library (MSCCL)~\cite{msccl}
on NVIDIA A100, NVIDIA H100, and AMD MI300x GPUs.
We show that collective communication kernels like AllReduce and AllGather implemented using the \mscclpp{} DSL are on average $1.99\times$, $2.08\times$, and $1.43\times$ faster than NCCL, RCCL, and MSCCL respectively. We also develop some of these communication algorithms directly using the \mscclpp{} Primitive API and show that those kernels are on average $3\%$ faster than their DSL counterparts.
These benefits translate to AI inference workloads. 
LLM inference decodes 
are on average $1.11\times$ and $1.31\times$ faster 
by using \mscclpp{} instead of NCCL 
in vLLM~\cite{kwon2023efficient} and SGLang~\cite{sglang} 
respectively.

The \mscclpp{} code base~\cite{mscclpp-github} has been publicly available for two years. During this time, RCCL~\cite{rccl} has adopted both the \mscclpp{} APIs and library as the default to support collective communication primitives on current and upcoming AMD hardware. 
SGLang~\cite{sglang}, a popular LLM inference framework, 
uses the \mscclpp{} library for its collective communication.
As a further validation of our design, when NVIDIA introduced support for \texttt{multimem} instructions for switch-mapped I/O (aggregation and multicast), incorporating this support in \mscclpp{}~(Section~\ref{sssec:switch-channel}) required only $16$ person-weeks, allowing the rest of the \mscclpp{} stack to leverage these primitives. Similarly, supporting NVIDIA's multi-node NVLink~\cite{mnnvl} took only $2$ person-weeks.

\section{Background and Motivation}
\label{sec:background}

\subsection{Collective Communications}
Communication in AI workloads often happens through collective operations in which multiple GPUs coordinate to transfer data among themselves and optionally performing computations on the data. 
\emph{AllReduce} is a common collective communication operation that sums the partial results of all GPUs and broadcasts the computed result to all GPUs. AllReduce can be divided into two other collective operations: \emph{ReduceScatter} and \emph{AllGather}. ReduceScatter sums the input buffers on all GPUs and distributes the output buffer equally on all GPUs. AllGather collects a distributed buffer from all GPUs and stores the full buffer on each GPU. 

\subsection{Vendor Communication Libraries}\label{ssec:nccl-limitations-2}
GPU vendors provide communication libraries that implement the collective operations.
NVIDIA Collective Communication Libraries (NCCL)~\cite{nccl} is an optimized library for collective communications on NVIDIA hardware.
Similarly, AMD ROCm Collective Communication Library (RCCL)~\cite{rccl} is optimized for AMD hardware.
Microsoft Collective Communication Library (MSCCL)~\cite{msccl} enables executing custom communication algorithm optimized for a particular input size and topology of the hardware. We briefly discuss the design of NCCL (see Hu~et~al.~\cite{hu2025demystifyingncclindepthanalysis} for a detailed description),
since both RCCL and MSCCL are based on NCCL and they both have the same limitations as NCCL.\footnote{RCCL has since adopted code and abstractions from \mscclpp{}.}

NCCL implements a GPU kernel for all collective operations. The implementation chooses the communication algorithms (ring, tree, etc.) to implement the collective based on the data size. The core communication happens through four primitive operations: \texttt{send}, \texttt{recv}, \texttt{copy}, and \texttt{reduce}.
NCCL allocates send and receive buffers to facilitate the transfer. These operations are synchronous. For instance, calls to \texttt{send} blocks until destination GPUs have called \texttt{recv} to ensure that it is safe to overwrite the send buffer.

\subsection{Limitations of NCCL Primitives}\label{sssec:nccl-limitations}
NCCL primitives are synchronous, which improves programmability, but wastes GPU cycles in busy-wait loops and prevents overlapping of computation and communication. In addition, NCCL's \texttt{send} and \texttt{recv} primitives work over internal send/receive buffers incurring local memory copies. Finally, by being a general-purpose library, NCCL hard codes many performance optimizations providing inflexibility to the users. For instance, NCCL provides limited capability to control the parallelism for performing reduction or copy operations during communication. Similarly, NCCL hardcodes a single transfer mode per link, even though interconnects like PCIe and NVLink support multiple modes of transfer with subtle performance tradeoffs.

\section{\mscclpp{} Overview}\label{sec:design}

\begin{figure}[t]
  \centering
  \includegraphics[width=\linewidth]{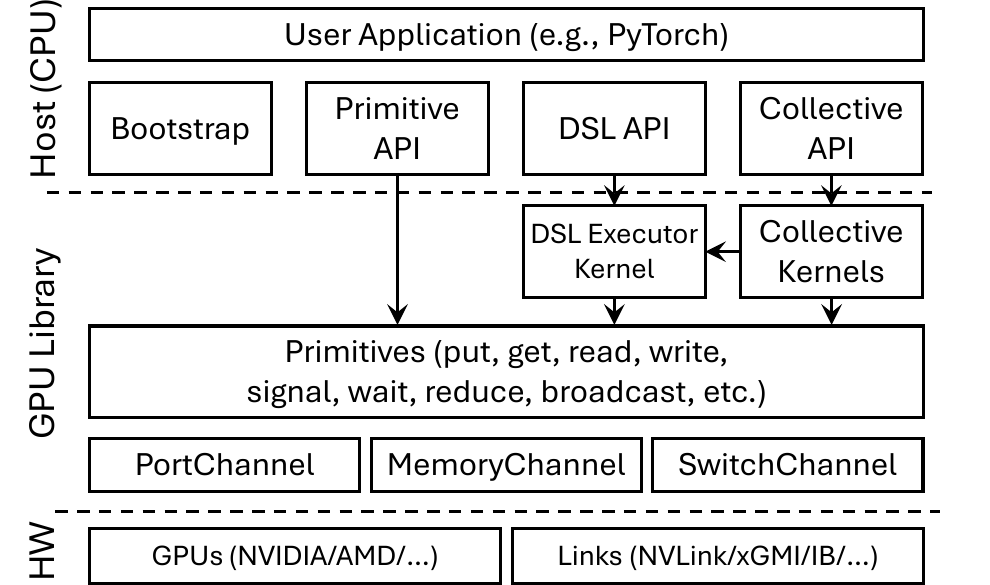}
  \caption{Overview of the \mscclpp{} stack.}
  \label{fig:mscclpp-overview}
  \vspace{-0.1in}
\end{figure}

\mscclpp{} provides three user APIs at hierarchical levels of communication abstractions: Primitive, DSL, and Collective APIs (see Figure~\ref{fig:mscclpp-overview}). The three levels provide different trade-offs in terms of programming effort or expertise and execution latency or performance. The closer the interface is to the hardware, the higher the programmer's control on execution, so higher the performance and programming effort.

\spara{\mscclpp{} Primitive API.} This replaces NCCL's primitives with a set of core communication functions that are called from the GPU kernels. This is paired with the bootstrapping API called from the host (CPU) side, which is used for connection setup between GPUs. The primitive interface is the base layer that is used to implement the other two higher-level interfaces. Programmers can also use the primitive interface directly to implement application-specific or hardware-specific optimizations in their own GPU kernels, including those kernels that also perform computation.

\spara{\mscclpp{} DSL API.}
Based on the primitive interface, we introduce a new DSL to write custom collective communication algorithms in a high-level language.
This interface is aimed at users who want to specify communication algorithms or schedules and generate kernels that are optimized for their own workloads (application and hardware).
The generated kernel uses the Primitive API by running the DSL Executor, so it may have small runtime overheads.

\spara{\mscclpp{} Collective API.} We re-implement the NCCL API, including its bootstrapping API, as-is over \mscclpp{}. It is aimed at users with the least expertise; they can simply replace NCCL/RCCL with the \mscclpp{} Collective library without changing their application code. Similar to NCCL/RCCL,
this library may not provide the best algorithm for certain workloads. 
For AI inference workloads, we implement the best algorithms in our collective kernels using the \mscclpp{} DSL.

\section{\mscclpp{} Primitive Abstractions}
\label{sec:primapi}

\begin{figure}[t]
    \begin{lstlisting}[language=C++,
    mathescape=true, numbers=left, numberstyle=\footnotesize,
    columns=fixed,
        basicstyle=\footnotesize\ttfamily,
        keywordstyle=\textcolor{blue},
        commentstyle=\color{commentgreen},
        firstnumber=1,
      escapeinside={(*}{*)}, % if you want to add comments in code
      morekeywords={blockDim, bid, class, warpSize, shared,
                    syncthreads, ulong, uint, bool},
              escapechar=|, xleftmargin=1.0ex,numbersep=4pt]
class PortChannel
  void *src, *dst; // set during init
  uint *semaphore; uint expectedVal;

  void put(ulong dstOff, ulong srcOff, ulong sz);
  void signal();
  void wait();
  void flush();

template<Protocol protocol> // LL or HB protocol
class MemoryChannel
  void *src, *dst; // set during init
  uint *semaphore; uint expectedVal;

  void put(ulong dstOff, ulong srcOff, ulong sz,
           uint tid, uint tids, uint flag=0);
  template<typename T>
  T read(ulong off, uint flag=0);
  template<typename T>
  void write(ulong off, T elem, uint flag=0);
  void signal();
  void wait();
  void flush();

class SwitchChannel
  void *local, *multimem; // set during init

  void reduce(ulong dstOff,ulong srcOff,ulong sz);
  void broadcast(ulong dstOff,ulong srcOff,
                 ulong sz);
\end{lstlisting}
\caption{\blue{Channel abstractions in \mscclpp{} Primitive API.}
}
\label{fig:primapi}
\end{figure}

\mscclpp{} Primitive API provides the minimal hardware abstractions for GPU communication.
These abstractions are close to the hardware to avoid performance overheads. At the same time, they hide the complexities of GPU/CPU/NIC orchestration, synchronization, and memory consistency issues to simplify the programmability. 

The modes of data transfer differ across different interconnects. At its core, \mscclpp{} provides the notion of a \texttt{Channel} to capture these modes of communication with basic synchronization primitives.
\blue{Specifically, \mscclpp{} provides}
(1) \texttt{PortChannel} for port-mapped I/O allowing GPUs to initiate data transfer through dedicated hardware I/O ports such as DMA engines or RDMA NICs, (2) \texttt{MemoryChannel} for memory-mapped I/O with direct peer-to-peer memory access across GPUs, and (3) \texttt{SwitchChannel} for switch-mapped I/O using \textit{multicasting} memory access enabled by interconnect switches. Each of these channels exposes additional primitives specific to that channel.

An interconnect might support multiple modes of data transfer and hence can support multiple of these interfaces. For instance, NVLink supports all the three, while xGMI and PCIe only support port-mapped and memory-mapped I/O. Our intent is that these abstractions are general allowing the library implementer to quickly adapt to changing hardware capabilities. For instance, the current implementation of \texttt{PortChannel} for NVLink 
requires coordination with the CPU to initiate a DMA engine (\texttt{cudaMemcpy}), but can directly initiate it from the GPU if future hardware supports it.

\blue{Port-mapped, memory-mapped, and switch-mapped I/O are complementary I/O methods in general computer architectures.
\texttt{PortChannel}, \texttt{MemoryChannel}, and \texttt{SwitchChannel} are our abstractions to provide these I/O methods from inside GPU kernels.
We believe that they are flexible enough to accommodate future hardware advances.
For example, if future hardware interconnects support initiating
DMA-copy from within a GPU kernel~\cite{ark}, the same \texttt{PortChannel} API
can target the new hardware.}

Figure~\ref{fig:primapi} shows three types of channels and its interface in \mscclpp{}.
\blue{\mscclpp{} also supports a few \textit{fused} primitives (omitted in Figure~\ref{fig:primapi} for brevity) that can reduce the overhead of API calls. For example, since a \texttt{put} is usually followed by a \texttt{signal}, we provide the \texttt{put\_with\_signal} function that performs both at once.
The buffers and semaphores are allocated on the GPU memory before initialization of the channels.}
Next, we describe the three kinds of \mscclpp{} channels.

\subsection{PortChannel}
\label{sec:portchannel}
A \texttt{PortChannel} implements primitives when data transfer is done over ports connected to GPU memory, such as \texttt{cudaMemcpy} for intra-node DMA or \texttt{ibv\_post\_send} for RDMA.
Borrowing the term from MPI~\cite{snir1998mpi}, writing data to the peer's side is called \textbf{\texttt{put}}. To provide our communication abstraction as close as possible to hardware capabilities, we design \texttt{put} to be zero-copy (i.e., no intermediate buffers), one-sided (i.e., initiated by a peer without participation of the other peer), and asynchronous.

Figure~\ref{fig:mscclpp-primitives} shows the semantics of the synchronization primitives: \texttt{signal}, \texttt{wait}, and \texttt{flush}. The \texttt{put} primitive transfers data from \texttt{src0} buffer of GPU-0 to \texttt{dst1} buffer of GPU-1. The \texttt{signal} is an asynchronous operation that is strictly ordered with respect to the previous transfer operations, such as \texttt{put} here. The matching \texttt{wait} primitive called by GPU-1 blocks till it receives GPU-0's \texttt{signal} at which time GPU-1 is allowed to access the transferred data in \texttt{dst1}. GPU-0 calls the \texttt{flush} primitive to ensure that the transfer is complete from its perspective allowing it to reuse the \texttt{src0} buffer. 

\spara{Implementation.}
Since data transfer over a port currently requires the CPU to initiate the transfer, each channel creates its own CPU thread.
The CPU thread reads data transfer or synchronization requests from a first-in-first-out request queue.
The storage and tail of a request queue are allocated using \texttt{cudaMallocManaged} as both CPU and GPU can access it, while the head is on GPU and only accessed by GPU.

\begin{figure}[t]
\centering
\includegraphics[width=0.8\linewidth]{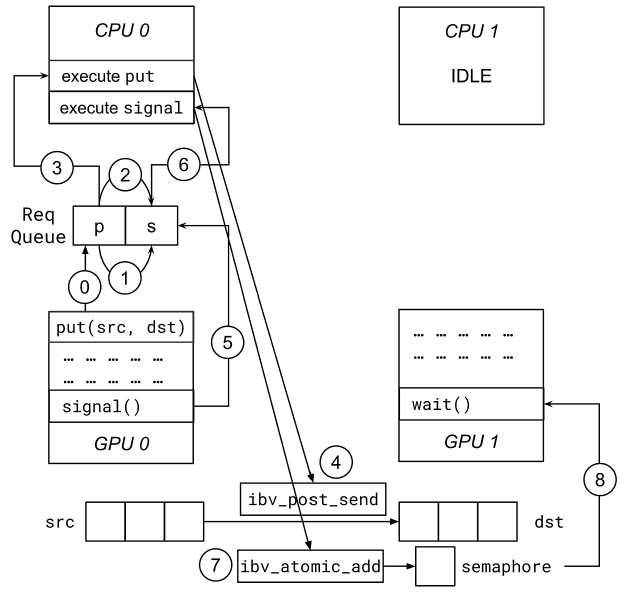}
\caption{\texttt{PortChannel} workflow for IB from \textcircled{0} when GPU 0 calls \texttt{put} primitive to \textcircled{8} when GPU 1 receives the data.}
\label{fig:mscclpp-ib}
\end{figure}

We now discuss the workflow of \texttt{PortChannel}s using Figure~\ref{fig:mscclpp-ib}.
\circled{0} When the GPU calls a primitive, such as \texttt{put}, the first participating thread of the GPU pushes the request to the queue by writing at the head.
Before writing to the queue, the GPU checks if the queue is filled, i.e., if the head value is more than the tail value.
If the queue is filled then the GPU waits for the CPU to process at least one request.
\circled{1} Then the GPU increments the head to the next element.
\circled{2} The CPU thread continuously reads the element at the tail to see if there is a request from the GPU, and when there is a request, the CPU reads the request, zeros out the current element, and increments the tail.
\circled{3} Now the CPU thread will process the request.
Below we explain
three types of requests
and how the CPU handles them over InfiniBand (IB) as an example.

\begin{figure}[t]
\begin{subfigure}[b]{0.47\columnwidth}
  \begin{lstlisting}[language=C,
    mathescape=true, numbers=none, numberstyle=\footnotesize,
    columns=fixed,
        basicstyle=\footnotesize\ttfamily,
        keywordstyle=\textcolor{blue},
        commentstyle=\color{commentgreen},
        firstnumber=1,
        rulecolor= \color{black},
        frame=single, %single line frame
        framerule=0.5pt, %thick frame
      escapeinside={(*}{*)}, % if you want to add comments in code
      morekeywords={blockDim, bid, tid, warpSize, shared,
                    syncthreads,put,signal,flush},
              escapechar=|, xleftmargin=1.0ex,numbersep=4pt]
//async
put(src0, dst1, size)
//unsafe to reuse src0
signal() //async
flush() //sync
//safe to reuse src0
\end{lstlisting}
\caption{GPU-0}
\end{subfigure}
\hfill
\begin{subfigure}[b]{0.44\columnwidth}
  \begin{lstlisting}[language=C,
    mathescape=true, numbers=none, numberstyle=\footnotesize,
    columns=fixed,
        basicstyle=\footnotesize\ttfamily,
        keywordstyle=\textcolor{blue},
        commentstyle=\color{commentgreen},
        firstnumber=1, showlines=true,
        rulecolor= \color{black},
        frame=single, %single line frame
        framerule=0.5pt, %thick frame
      escapeinside={(*}{*)}, % if you want to add comments in code
      morekeywords={blockDim, bid, tid, warpSize, shared,
                    syncthreads,wait},
              escapechar=|, xleftmargin=1.0ex,numbersep=4pt]

//unsafe to read dst1
wait() //sync
//safe to read dst1

\end{lstlisting}
\caption{GPU-1}
\end{subfigure}

  \caption{\mscclpp{} data transfer abstractions.
\texttt{put} asynchronously transfers data from one GPU to another.
\texttt{signal} and \texttt{wait} synchronize data transfer between GPUs.
\texttt{flush} ensures the completion of preceded data transfer.}
  \label{fig:mscclpp-primitives}
\end{figure}

\spara{Data Transfer.} \texttt{put} pushes a put request in the queue.
\circled{4} The CPU processes this request by starting an RDMA transfer using \texttt{ibv\_post\_send}.
Since this function is asynchronous, the CPU thread immediately returns.
When the transfer is happening, peer-GPUs are free to execute code, thereby, improving the overall power efficiency of the system.

\spara{Synchronization.} \circled{5} \texttt{signal} pushes a signal request in the queue.
\circled{6}\circled{7} The CPU processes this request by atomically incrementing the semaphore on the receiving GPU using functions, like \texttt{ibv\_atomic\_add} for IB.
The \texttt{wait} primitive on the receiving GPU does not create a request for its CPU, therefore, the receiving CPU is idle, rather \texttt{wait} continuously looks for the semaphore to reach an expected value in a while loop.
\circled{8} \texttt{wait} returns after the semaphore is incremented.

\spara{Flush.} Primitive \texttt{flush} pushes a flush request in the queue and the first thread of the GPU waits until the queue tail reaches the position where the flush request is pushed to.
The CPU processes the flush request by waiting until all the previous data transfer and synchronization requests have been completed.
For example, for IB we use \texttt{ibv\_poll\_cq} to get the status of all requests.
After the flush request is complete, the GPU is free to re-write to the source buffer.

\subsection{MemoryChannel}
\label{subsec:memory-channel}
A \texttt{MemoryChannel} wraps data transfer methods that use thread-copy mechanism, i.e., directly use GPU threads for writing to peer GPU's memory.

\spara{Protocols.}
The channel provides two protocols, which define data transfer and synchronization technique to tune between low-latency and high-bandwidth:
\emph{HB} protocol provides a high-bandwidth but high-latency protocol, thus, is suitable for larger sizes, and
\emph{LL} protocol provides low-latency but low-bandwidth, thus, is suitable for smaller sizes.
Both protocols achieve these properties by synchronizing on the suitable granularity of data.

(i) In the HB protocol, peer GPUs transfer a large chunk of data and synchronize this chunk once using wait and signal primitives.
Hence, the synchronization time is amortized over the transfer time leading to high bandwidth.
However, since the receiving GPU needs to wait for the whole chunk before processing it, the HB protocol also has high latency.

(ii) In LL protocol, the receiving GPU synchronizes on fewer chunks of data being transferred and processes the chunk as soon as it is received, thus achieving lower latency.
However, since the number of synchronization is proportional to the number of elements, this protocol provides lower-bandwidth.
The LL protocol works as follows.
The \texttt{put} primitive requires an extra integer \texttt{flag} value.
For every $N-1$ elements written to the receiving GPU, \texttt{put} also writes the \texttt{flag}.
The receiving GPU uses the \texttt{read} primitive, waits until the \texttt{flag} value at $N$ index of receiving buffer is set, and then reads and returns the $N-1$ elements.
However, we cannot arbitrarily use any $N$ because GPUs follow a weak memory consistency model in which writes to different memory locations by multiple threads can be performed in any order.
Therefore, we restrict $N$ to number of elements written by a single instruction, i.e., 4, 8, and 16 bytes memory accesses instructions.
The \texttt{flag} value is decided based on the algorithm, such that, all \texttt{flag} values are distinct.

\spara{Data Transfer.} \texttt{put} is a collective-thread operations that reads data elements from source buffer and writes to destination buffer starting from different offsets using a set of threads.
Similarly, the \texttt{read} primitive reads and returns data from the destination buffer, and the \texttt{write} primitive writes data including the \texttt{flag} for the LL protocol to the destination buffer.
To maximize bandwidth for the HB protocol, both primitives use 16-bytes loads and stores.
For the LL protocol, both primitives also take a flag and use 8-byte loads and stores by default or user supplied vector length.
Since both \texttt{put} primitives are called by multiple threads (all or first few threads of the kernel), we achieve maximum bandwidth.

\spara{Synchronization.} Synchronization primitives wait on a semaphore that is an integer allocated on the receiving GPU.
The \texttt{signal} primitive atomically increments the receiving GPU's semaphore and calls \texttt{threadfence\_system} to ensure that writes by \texttt{put} and semaphore increments are made available in this order.
The \texttt{wait} primitive performs a busy-wait while-loop that checks until the value of semaphore has reached the expected value.
This wait is performed by the first thread of the kernel and all other threads wait on a kernel barrier.
The channel tracks the expected value using its \texttt{expectedValue} member.
The \texttt{flush} primitive is a no-op because after \texttt{put} returns, the source buffer can be reused, even though the write is still in progress.

\subsection{SwitchChannel}\label{sssec:switch-channel}
A \texttt{SwitchChannel} provides two primitives for performing collective operations among GPUs:
\texttt{reduce} to add the corresponding elements of buffers residing on different GPUs and \texttt{broadcast} to send elements from a buffer on a GPU to all other GPUs.
These operations usually require specialized hardware support.
For example, NVIDIA NVLink 4.0 connects all H100 GPUs to a single NVSwitch and this NVSwitch can perform
in-network, switch-based aggregation and multicast using NVLink SHARP (NVLS) technology~\cite{nvls}.
\blue{
We now discuss the implementation of these primitives for NVSwitch, but
we believe that these implementations can be generalized to other hardware in the future.
}

\spara{Reduce.} This primitive takes a destination buffer as a local address on the local GPU and a source buffer allocated as a \emph{multimem} address.
A multimem address is a virtual address that points to different virtual addresses on each GPU that is a part of the channel/collective.
The primitive goes through each element of the destination buffer on the local GPU and executes \texttt{multimem.ld\_reduce} PTX instruction 
using the source element's multimem address. 
The multimem instruction 
fetches the values from all the virtual memory addresses pointed by the multimem address to the switch, does the reduction on the switch, and returns the reduced value to the local GPU.
The reduced output is obtained in a register and then written to the destination buffer.

\spara{Broadcast.} This primitive takes a source buffer as a local address on the local GPU and a destination buffer as a multimem address.
The primitive goes through each element of the source buffer on the local GPU, reads the element into a register, and executes a \texttt{multimem.st} PTX instruction using the source register and the destination element's multimem address. 
The multimem instruction sends the register value to the switch, which broadcasts and stores the value to all virtual memory addresses pointed by the multimem address.

\subsection{Advantages of Primitive API}
Expert developers can use the Primitive API to build custom workload-specific communication algorithms in a portable manner \blue{with minimal performance overhead (see Table~\ref{tbl:primitive-perf})}. At the same time, developers implementing these primitives can add support for emerging hardware and interconnects. 
By separating primitives for data transfer and synchronization, \mscclpp{} allows asynchronous communication and batching synchronization of multiple data transfers.
By using channels for specific GPU interconnects, \mscclpp{} allows users to specialize kernels for particular interconnects.
Furthermore, \mscclpp{} avoids extra copies in kernels, leading to a kernel with less code paths and zero register spills.
Since the Primitive API is an in-kernel interface for communication, we can co-optimize both computation and the communication kernels, leading to better application performance.
One example is overlapping reduction in AllReduce 
with communication
\blue{(see Section~\ref{ssec:overlapped-ring-rs}).}

\begin{table}[t]
\small
\begin{tabular}{|c|c|c|c|}
\hline
                              & Best Achievable & \mscclpp{} \\ \hline
NVLink Throughput (GB/s)      & 397.5           & 397.5   \\
NVLink Latency (ns)           & 822             & 829     \\ \hline
InfiniBand Throughput (GB/s)  & 48.94           & 48.94   \\
InfiniBand Latency (\micro s) & 3.76            & 4.89    \\
\hline
\end{tabular}
\caption{\blue{\mscclpp{} Primitive API's peer-to-peer performance compared to the best achievable on our H100 environment (see Table~\ref{tbl:env}). The best achievable performance is measured by \texttt{nvbandwidth}~\cite{nvbandwidth} and RDMA \texttt{perftest}~\cite{perftest}. \mscclpp{}'s NVLink throughput and latency are measured with \texttt{PortChannel} and \texttt{MemoryChannel}, respectively. \mscclpp{}'s InfiniBand throughput and latency are measured with \texttt{PortChannel}.}}
\label{tbl:primitive-perf}
\end{table}

\section{\mscclpp{} DSL}\label{ssec:dsl-impl}

\begin{figure}[t]
    \begin{lstlisting}[
    mathescape=true, numbers=left, numberstyle=\footnotesize,
    columns=fixed,
        basicstyle=\footnotesize\ttfamily,
        keywordstyle=\textcolor{blue},
        commentstyle=\color{commentgreen},
        firstnumber=1,
      escapeinside={(*}{*)}, % if you want to add comments in code
      morekeywords={def,class},
      escapechar=|, xleftmargin=1.0ex,numbersep=4pt]
class PortChannel:
    def __init__(self, dst_rank: int, 
        src_rank: int): ...
    def put(self, dst_chunk: Buffer, 
        src_chunk: Buffer, tb: int): ...
    def signal(self, tb: int): ...
    def wait(self, tb: int): ...
    def flush(self, tb: int): ...

class MemoryChannel:
    def __init__(self, dst_rank: int, 
        src_rank: int): ...
    def put(self, dst_chunk: Buffer, 
        src_chunk: Buffer, tb: int, 
        tb_group: ThreadBlockGroup): ...
    def put_packets(self, dst_chunk: Buffer, 
        src_chunk: Buffer, tb: int, 
        tb_group: ThreadBlockGroup): ...
    def reduce(self, dst_chunk: Buffer, 
        src_chunk: Buffer, reduce_op: ReduceOp,
        tb: int, tb_group: ThreadBlockGroup): ...
    def signal(self, tb: int): ...
    def wait(self, tb: int): ...

class SwitchChannel:
    def __init__(self, ranks: List[int]): ...
    def broadcast(self, 
        dst_multimem_chunk: MultimemBuffer, 
        src_chunk: Buffer, tb: int): ...
    def reduce(self, dst_chunk: Buffer, 
        src_multimem_chunk: MultimemBuffer, 
        reduce_op: ReduceOp, tb: int): ...
\end{lstlisting}
\caption{\blue{Channel abstractions in \mscclpp{} DSL (Python).}
}
\label{fig:dsl-api}
\end{figure}

To enable specifying custom communication algorithms, \mscclpp{} supports a Python-based Domain-Specific Language (DSL) and lowers the algorithm into a sequence of \blue{operations} that can be executed by the DSL Executor on top of the primitive abstractions described above. 
The primitives exposed at the DSL retain the zero-copy, one-sided, and asynchronous properties of the underlying primitive abstractions, thereby enabling users to specify optimized algorithms that overlap compute and communication.

\begin{figure}[t]
\begin{lstlisting}[language=Python,
mathescape=true, numbers=left, numberstyle=\footnotesize,
columns=fixed,
    basicstyle=\footnotesize\ttfamily,
    keywordstyle=\textcolor{blue},
    commentstyle=\color{commentgreen},
    firstnumber=1,
  escapeinside={(*}{*)}, % if you want to add comments in code
  escapechar=|, xleftmargin=1.0ex,numbersep=4pt]
def ringRS(portChannels, bufs, sz, N, tb=0):
  scr = [Buffer(rank, sz) for rank in range(N)]
  for rank in range(N):
    putChan   = portChannels[(rank+1)%N]
    recvChan  = portChannels[(rank+N-1)%N]
    src       = bufs[rank]
    recv, dst = scr[rank], scr[(rank+1)%N]

    def chunk(step):
      csz = sz / N
      offset = ((rank + N - step) % N) * csz
      return offset, offset+csz/2, offset+csz

    for step in range(N): 
      beg, mid, end = chunk(step)
      # (a) Put 1st half of chunk
      putChan.put(dst[beg:mid], src[beg:mid], tb) |\label{line:put-first-half}|
      putChan.signal(tb)
      if step != 0: 
        pbeg, pmid, pend = chunk(step-1)
        # Reduce prev 2nd half, overlaps (a) 
        src[pmid:pend].reduce(recv[pmid:pend], tb) |\label{line:reduce-second-half}|
      recvChan.wait(tb) # Wait for 1st half
      putChan.flush(tb) # Flush 1st half
      # (b) Put 2nd half of chunk
      putChan.put(dst[mid:end], src[mid:end], tb) |\label{line:put-second-half}|
      putChan.signal(tb)
      # Reduce 1st half, overlaps (b)
      src[beg:mid].reduce(recv[beg:mid], tb)  |\label{line:reduce-first-half}|
      recvChan.wait(tb) # Wait for 2nd half
      putChan.flush(tb) # Flush 2nd half
      if step == N-1:
        src[mid:end].reduce(recv[mid:end], tb)

\end{lstlisting}
\caption{A simplified example of a Ring ReduceScatter implementation using \mscclpp{} DSL.
\label{fig:ring-reduce-scatter-dsl}}
\end{figure}

\subsection{DSL}

\mscclpp{} DSL provides a global view of all thread blocks on all GPUs or ranks in a Python-native language. 
\blue{As shown in Figure~\ref{fig:dsl-api},}
the DSL exposes \texttt{PortChannel}, \texttt{MemoryChannel}, and \texttt{SwitchChannel} abstractions in this global view. 
Consequently, unlike the Primitive API (Section~\ref{sec:primapi}), users specify the ranks involved in each channel and for each \blue{operation} in the channel, specify the thread block on the channel's source rank that must execute it. 
\blue{Some operations on the \texttt{MemoryChannel} also accept an optional \texttt{ThreadBlockGroup} that enables a group of thread blocks to collectively perform that operation.
The \texttt{Channel} classes include a few \textit{fused} primitives (omitted in Figure~\ref{fig:dsl-api} for brevity) that can reduce the overhead of API calls. For example, \texttt{MemoryChannel}'s
\texttt{reduce\_put} \texttt{reduce}s two local buffers and \texttt{put}s the result to a remote destination using a temporary register to avoid a memory round-trip for the intermediate data.}
In addition, the DSL provides a separate \texttt{Buffer} abstraction and the \texttt{Channel}s are not tied to a pair of buffers, so users specify the chunks of the source and destination buffer in each data transfer \blue{operations}. 
Since the DSL is Python-native, the chunks are specified as slices of \texttt{Buffer}. \texttt{Buffer}s are associated with a rank and allow computation like \texttt{reduce}, which is performed on its rank. 
The DSL also provides convenience classes to create input and output buffers based on the communication pattern 
(e.g., AllReduce, AllGather, ReduceScatter), 
device-wide synchronization primitives, and multi-device 
synchronizations primitives; synchronization within a thread block is handled automatically by the DSL.
\blue{Users can control chunk sizes, number of queue pairs (QPs), etc. via the DSL. For instance, if users want two QPs, they can use two parallel \texttt{PortChannel}s. If users want to offload the communication to CPU, they can use \texttt{PortChannel} rather than \texttt{MemoryChannel}.}
Thus, the \mscclpp{} DSL enables users to construct complex data movement and synchronization workflows without writing low-level CUDA code.

\subsection{Example: Overlapped Ring ReduceScatter}\label{ssec:overlapped-ring-rs}
Figure~\ref{fig:ring-reduce-scatter-dsl} demonstrates how the DSL can implement an efficient Ring ReduceScatter algorithm that overlaps computation and communication.
In the classic Ring algorithm, $N$ GPUs are organized in a ring, the input buffer is divided into $N$ chunks, and each chunk circulates around the ring to be reduced. At each step, a GPU receives a chunk, reduces it locally, and then forwards it to the next GPU, with all steps occurring synchronously. To enable overlap, Figure~\ref{fig:ring-reduce-scatter-dsl} further splits each chunk into two halves. This allows the reduction of the first half (Line~\ref{line:reduce-first-half}) to overlap with the communication of the second half (Line~\ref{line:put-second-half}), and the reduction of the second half (Line~\ref{line:reduce-second-half}) to overlap with the communication of the first half (Line~\ref{line:put-first-half}).
In Figure~\ref{fig:ring-reduce-scatter-dsl}, 
only thread block 0 on each GPU does this computation and communication. 
The user can experiment with more parallelism or pipelining by using more thread blocks on each GPU and/or using more chunks per GPU.
The \mscclpp{} DSL supports replicating program instances to increase parallelism and improve performance 
while automatically handling mapping of the program 
to thread blocks and channels.

\extended{
While \mscclpp{} DSL raises the level of abstraction, 
it also introduces some restrictions compared to the Primitive API.
A DSL program is restricted to one protocol 
(Section~\ref{subsec:memory-channel}): either LL or HB; 
that is, all channels within a program must use the same protocol. 
\texttt{MemoryChannel}'s \texttt{put} can only be executed by one thread block (unlike in Figure~\ref{fig:primapi}). Users can parallelize \texttt{put} across multiple thread blocks by making multiple calls and adding explicit device-wide synchronization if necessary. 
These restrictions are similar to those in prior DSLs like 
MSCCLang~\cite{cowan2023mscclang}. However, unlike MSCCLang, 
\mscclpp{} DSL enables a single GPU thread block to access multiple GPUs at the same time. 
}

\subsection{Lowering and Optimizations}\label{ssec:dsl-opt}
The program written in \mscclpp{} DSL is lowered to an \textit{execution plan}, which consists of information from the program such as the collective, channel types, data transfer protocol, memory buffers to register for each thread block, semaphores to setup, and the sequence of operations \blue{(which may include loops)} to run in each thread block of every rank. 
\blue{Buffer sizes and rank counts (\texttt{N} in Figure~\ref{fig:ring-reduce-scatter-dsl}) are provided by the user at lowering time.}
During lowering, the program goes through data dependence analysis to add appropriate synchronizations, as well as optimization passes such as \blue{operation} fusion.
This enables the DSL to achieve performance comparable to hand-tuned CUDA implementations.

\spara{Data Dependence Analysis.} \mscclpp{} DSL automatically tracks data dependencies at the chunk level within each thread block by maintaining the last writer and active readers for each memory slot. When operations have data dependencies, the lowered program includes necessary synchronization points to ensure correct execution order. 
For example, in Figure~\ref{fig:ring-reduce-scatter-dsl}, 
it introduces a thread block synchronization on thread block 0 before the reduce operation 
(Line~\ref{line:reduce-first-half}).
\blue{The DSL also detects redundant synchronizations and removes them. Specifically, if a lowered program will be calling multiple thread block synchronizations back-to-back, the redundancies will be removed, retaining only one of them.}

\spara{\blue{Operation} Fusion.} \mscclpp{} DSL also maintains a directed acyclic graph (DAG) to track data dependencies and usage patterns of operations in a thread block at the chunk level. When two or more operations meet the fusion criteria, such as contiguous chunk access, no intervening dependencies, and compatible resource requirements, the DSL merges them into a single operation function. This fusion strategy reduces memory traffic and avoids unnecessary synchronization, resulting in faster execution.
\blue{For example, a DSL code using \texttt{MemoryChannel}:}
\begin{Verbatim}[fontsize=\small]
    src.reduce(data,tb); memChan.put(dst,src,tb);
\end{Verbatim}
\blue{is captured into a fused operation:}
\begin{Verbatim}[fontsize=\small]
    memChan.reduce_put(dst,src,data,tb);
\end{Verbatim}
\blue{where \texttt{reduce\_put} \texttt{reduce}s two local buffers and \texttt{put}s the result to a remote destination using a temporary register to avoid a memory round-trip for the intermediate data.}

\begin{table*}[t]
    \center\small
    \begin{tabular}{|c|c|c|c|}
    \hline
    Env. Name & GPU & Intra-node Link & Network \\
    \hline
    A100-40G & NVIDIA A100 (40G) (8x/node) & NVLink 3.0 & Mellanox HDR InfiniBand (200 Gb/s, 1x NIC/GPU) \\
    A100-80G & NVIDIA A100 (80G) (8x/node) & NVLink 3.0 & Mellanox HDR InfiniBand (200 Gb/s, 1x NIC/GPU) \\
    H100 & NVIDIA H100 (8x/node) & NVLink 4.0 & Quantum-2 CX7 InfiniBand (400 Gb/s, 1x NIC/GPU) \\
    MI300x & AMD MI300x (8x/node) & Infinity Fabric Gen 4 & Quantum-2 CX7 InfiniBand (400 Gb/s, 1x NIC/GPU) \\
    \hline
    \end{tabular}
    \caption{List of environments used for evaluation.}
    \label{tbl:env}
    \vspace{-0.1in}
\end{table*}

\subsection{Executor}
Given an execution plan and the input and output buffers, the DSL Executor initializes different channels, setting up connections and semaphores, and registering memory. Finally, the executor runs an execution kernel that interprets the list of operations in the execution plan.
\blue{The execution kernel is a single GPU kernel implementation that runs any given execution plan.}
The execution kernel inlines calls to primitive \mscclpp{} operations such as \texttt{put}, \texttt{signal}, \texttt{wait}, and \texttt{flush}. The execution plan can also contain fused operations, such as \texttt{ReduceSend} explained in Section~\ref{ssec:dsl-opt}.

\section{\mscclpp{} Collectives Library}\label{sec:algos}
\mscclpp{} Collective API is the simplest API that \mscclpp{} offers, and it is the same as the popular NCCL or RCCL API. We implement this API using kernels written in the \mscclpp{} DSL. Users can either use these default kernels or plug-in their own algorithms written using the \mscclpp{} DSL or Primitive APIs.
We now briefly describe a few AllReduce algorithms included in the default kernels to: 
(i) provide ideas on low-level considerations for implementing high-performance GPU communication kernels, and 
(ii) to show how low-level interfaces can be useful for customization on different workloads.

\spara{1. One-phase All-pairs (1PA).} In \textit{all-pairs} algorithms, all GPUs concurrently broadcast their own local data to all other GPUs. By \textit{one-phase}, the algorithm sends all local data to all other GPUs, so that the reduction is done in a single phase. It is suitable for small message sizes where the synchronization overhead between GPUs is more critical than the redundant reduction and data traffic. As we use the 1PA algorithm only for very small message collectives within a single node, we implement only a single version of 1PA that uses the LL protocol with \texttt{MemoryChannel}.
\blue{\mscclpp{} can implement it much more efficiently than other libraries by 
relaxing synchronizations 
and concurrently transferring data to multiple devices 
before waiting on any of them.}

\spara{2. Two-phase All-pairs (2PA).} By \textit{two-phase}, the algorithm splits the AllReduce into two phases: the first for ReduceScatter (each of $N$ GPUs collects and reduces $1/N$ of the data) and the second for AllGather (each GPU broadcasts the reduced data to all other GPUs)~\cite{Rabenseifner}. Two-phase algorithms are more bandwidth-efficient and conduct less reduction than one-phase algorithms. In 2PA, the ReduceScatter and AllGather phases are done in the all-pairs manner each.
We use 2PA for single-node collectives and implement multiple variants using \texttt{PortChannel},
\texttt{MemoryChannel} with LL or HB protocol, or \texttt{SwitchChannel}.
\blue{Using \mscclpp{}, we can optimize 2PA in various ways that are not possible in existing libraries. For example, for up to a few MBs of messages, we exploit rotating buffers to reduce synchronization at the cost of using more memory space.
As another example, we can let a single thread group read data from multiple other GPUs at the same time. This allows efficient data reduction compared with other libraries that read data from different GPUs one-by-one, which synchronizes at each reduction step.}

\spara{3. Two-phase Ring (2PR).} Similar to 2PA, 2PR has two phases: 
the first phase does ReduceScatter in a ring 
and the second phase does AllGather in the same ring. 
\blue{Unlike NCCL, we can use \texttt{PortChannel} (DMA-copy) 
even within a node and we overlap the reduction with the DMA-copy. Figure~\ref{fig:ring-reduce-scatter-dsl} 
shows a simplified kernel for ReduceScatter. We extend 
such pipelining to overlap reduction with communication 
across both phases.
Our 2PR implementation using \texttt{PortChannel} shows the best throughput among all implementations for intra-node AllReduce with large message sizes.}

\spara{4. Two-phase Hierarchical (2PH).} Hierarchical algorithms exchange minimal data across nodes and do local collectives in each node to complete the operation. It can be faster than all-pairs by reducing the data traffic crossing the nodes. 2PH is a two-phase algorithm that performs ReduceScatter and AllGather in a hierarchical manner each. We use it for multi-node collectives and implement two variants. The first version is for small messages using the LL protocol. Each node conducts local ReduceScatter that splits the data into the number of GPUs in a node. This requires sending more data across nodes and introduces redundant reduction, but it is faster for small messages by reducing synchronization steps. Cross-node communication is done in all-pairs manner. To utilize inter-/intra-node links at the same time, we pipeline the local collective with cross-node communication to overlap. The second version of 2PH is for large messages using HB protocol. This variant also performs all-pairs cross-node communication and local ReduceScatter in a pipelined manner.
Unlike the first variant, to utilize the link bandwidth efficiently, the number of data chunks is the same as the number of GPUs.

\section{Evaluation}
\label{sec:eval}

\spara{Environments.} Table~\ref{tbl:env} lists the environments used in the evaluation. Each node is equipped with 8 GPUs per node (either A100, H100, or MI300x) with intra-node links between GPUs (either NVLink or Infinity Fabric (a.k.a. xGMI)) and inter-node IB links (one NIC per GPU). All NICs are connected to a single IB networking switch. NVIDIA GPUs use CUDA 12.4, while AMD GPUs use ROCm 6.2. For brevity, the rest of this paper refers to each environment by the name of GPU.

\spara{Baselines.} We compare \mscclpp{} with the state-of-the-art collective communication libraries, including NCCL 2.26.2~\cite{nccl}, RCCL 2.20.5~\cite{rccl},\footnotetrim{As RCCL has adopted \mscclpp{}, we refer compare RCCL with \mscclpp{} disabled.} and MSCCL 2.23~\cite{msccl-github}. 

\begin{figure}[t]
    \centering
    \includegraphics[width=\columnwidth]{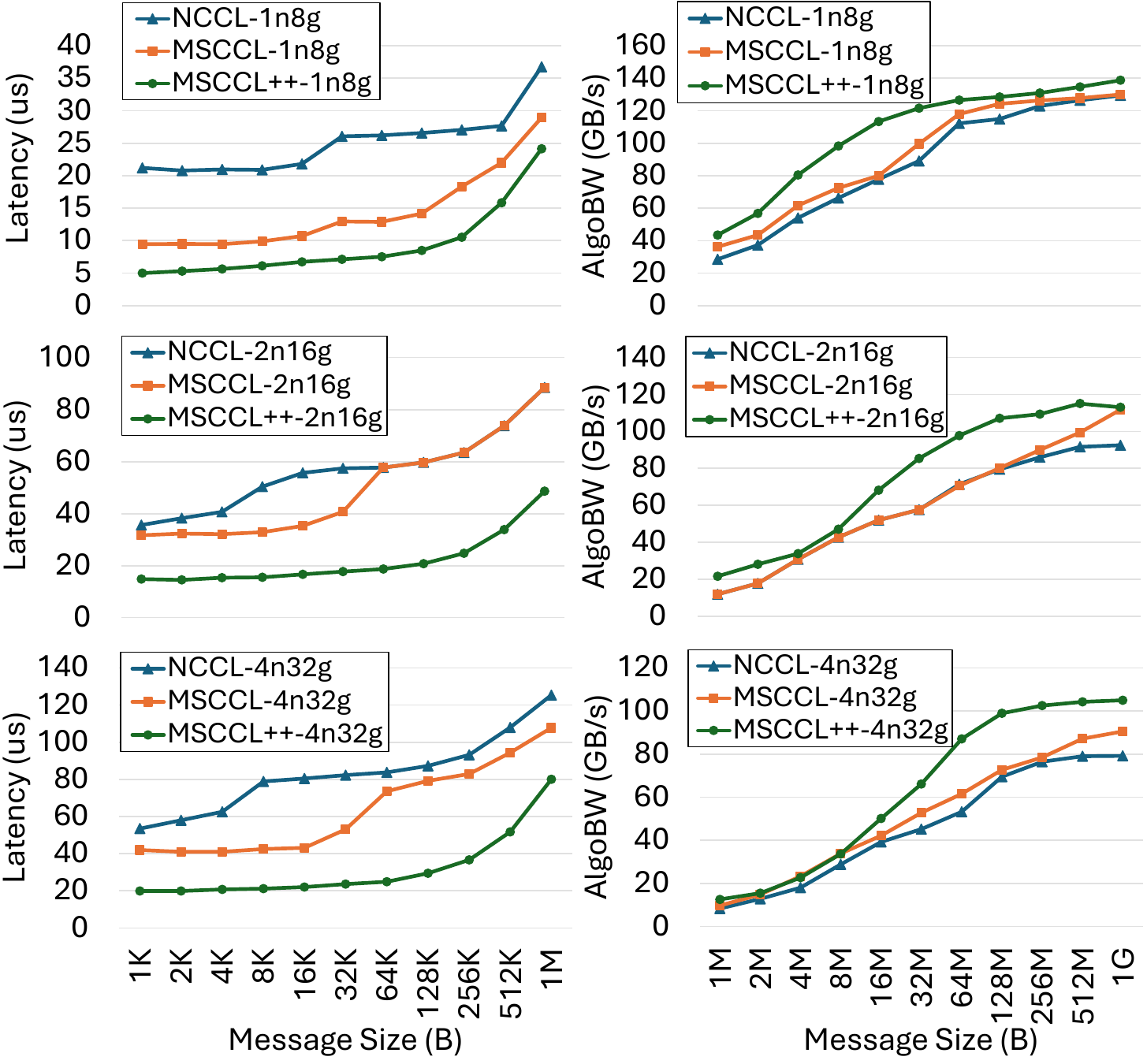}
    \caption{AllReduce, A100-40G. \textit{M}n\textit{N}g means \textit{M} nodes and \textit{N} GPUs.
    X-axis is message size (at the bottom of the figure).}
    \label{fig:all-reduce-a100}
\end{figure}

\subsection{Collective Communication Results}\label{sssec:ccl-perf}
To compare the performance of collective communication, we implement the AllReduce algorithms described in Section~\ref{sec:algos} using the \mscclpp{} DSL. We also implement various AllGather algorithms in a similar manner, and we omit the implementation details. We present the best number among all implementations for each message size and environment\footnote{\blue{We did not do sophisticated autotuning beyond picking the best configuration for a given platform and input size ranges (akin to MSCCLang~\cite{cowan2023mscclang}) by offline profiling. Our experiments measure the performance of the resulting system. Orthogonal techniques from TACCL~\cite{taccl} and TE-CCL~\cite{te-ccl} can be used for automation, which is an interesting future direction.}}.
All NCCL, RCCL, and MSCCL numbers are fine-tuned for each environment and message size by adjusting their environment variables, such as the number of channels (affects the number of threads), chunk size (affects the size of data to be transferred at once), type of algorithm (such as ring, tree, or NVLS~\cite{nvls}), the topology (XML file that describes the intra-node link topology of GPUs), etc. For MSCCL, we use the fastest algorithm for each message size~\cite{msccl-scheduler-github}. We use NCCL's user buffer registration API (i.e., \texttt{ncclMemAlloc})~\cite{nccl-buf-reg} and the CUDA/HIP Graph APIs for best performance.

For clarity of exposition, we separate message sizes into small (up to 1MB, presented as latency) and large (1MB and above, presented as algorithm bandwidth (AlgoBW)\footnotetrim{Defined as the message size divided by the latency.}) in the figures. The small message sizes represent inference scenarios (such as LLM token sampling a.k.a. \textit{decode}~\cite{scaling-transformer}), while large message sizes represent both training and inference scenarios (such as gradient accumulation during back-propagation, or LLM prompt processing a.k.a. \textit{prefill}~\cite{scaling-transformer}).

\begin{figure}[t]
    \centering
    \ifmscclpp
    \includegraphics[width=\columnwidth]{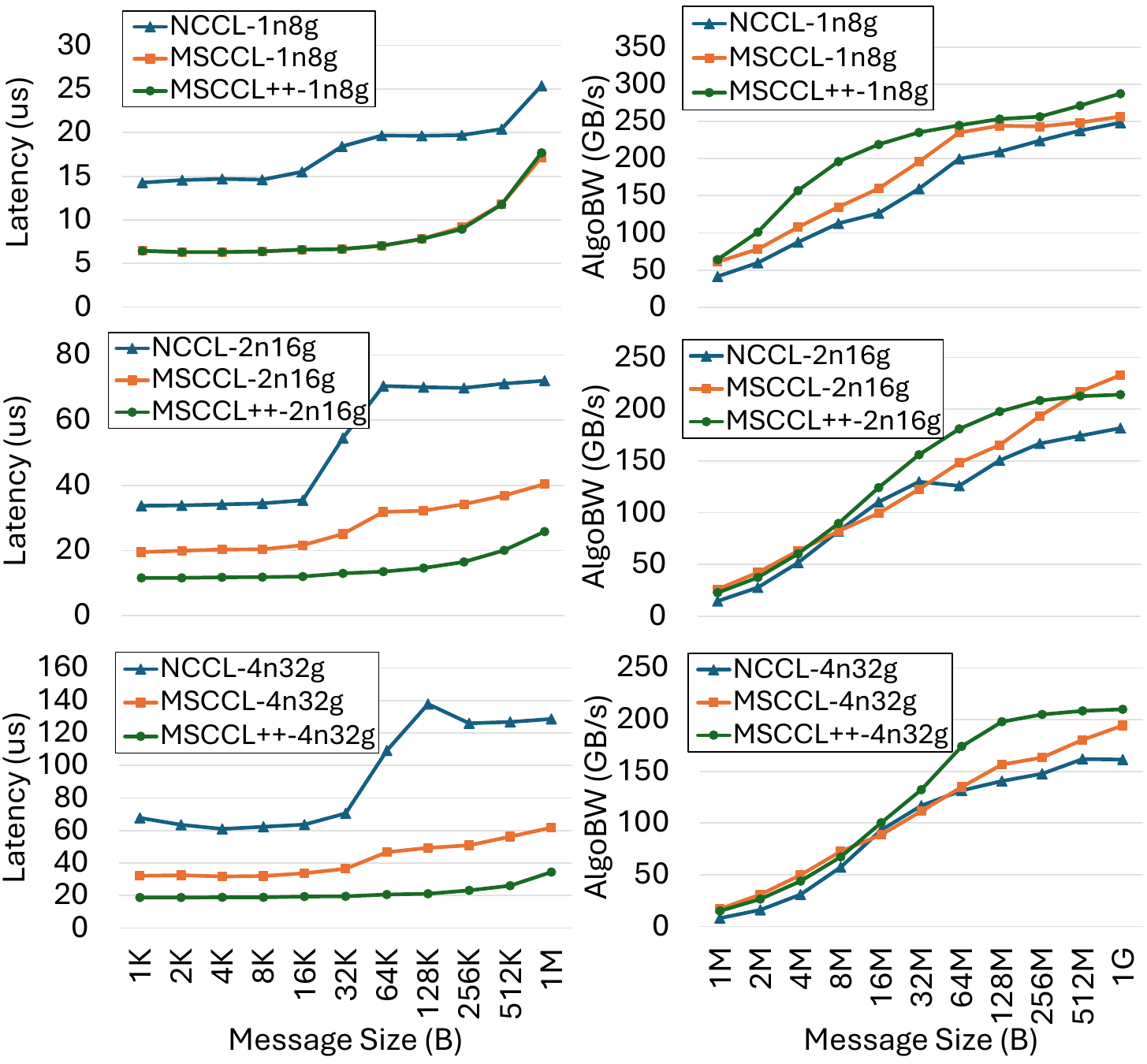}
    \else
    \includegraphics[width=\columnwidth]{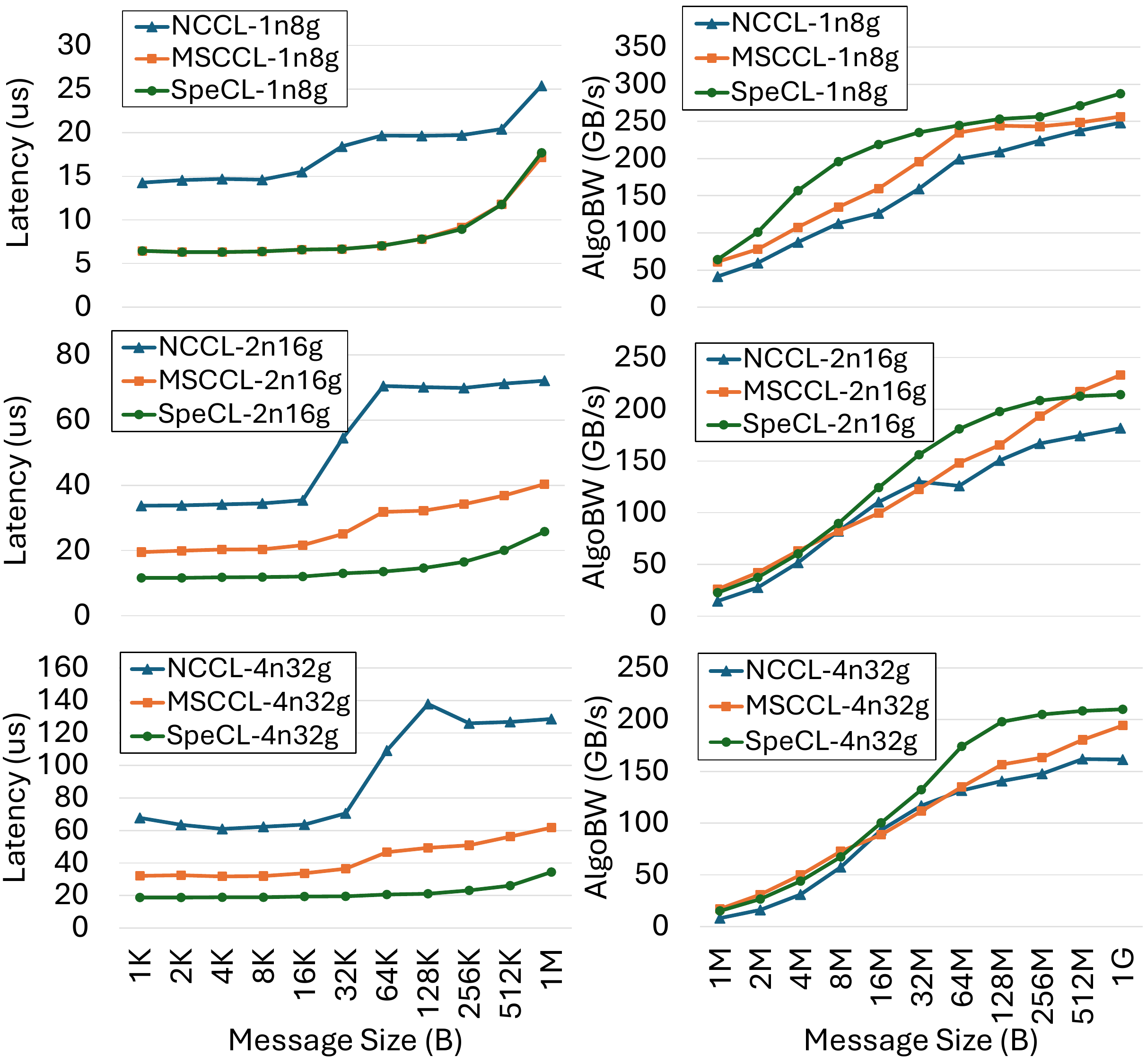}
    \fi
    \caption{AllGather, A100-40G. \textit{M}n\textit{N}g means \textit{M} nodes and \textit{N} GPUs.
    X-axis is message size (at the bottom of the figure).}
    \label{fig:all-gather-a100}
\end{figure}

\spara{Results.} Figure~\ref{fig:all-reduce-a100} compares the AllReduce performance over A100-40G. \mscclpp{} outperforms baselines for both small messages (up to \textbf{4.2x} and \textbf{3.1x} faster over NCCL and MSCCL, respectively) and large messages (up to \textbf{1.8x} faster over both NCCL and MSCCL). Similarly, Figure~\ref{fig:all-gather-a100} compares the AllGather performance over A100-40G. For small messages, \mscclpp{} outperforms NCCL and MSCCL by up to \textbf{5.4x} and \textbf{2.3x}, respectively, and for large messages, \mscclpp{} outperforms NCCL and MSCCL by up to \textbf{1.8x} and \textbf{1.4x}, respectively. \mscclpp{} is faster than other baselines across the board for AllReduce. For AllGather, MSCCL outperforms \mscclpp{} by 8.0\% in one scenario. This is due to the performance overhead of our DSL API that we are able to reduce if we implement directly on top of \mscclpp{}'s Primitive API.  

\spara{Gain Breakdown.} 
The performance gain of \mscclpp{} comes from both algorithmic improvements, fine-grained overlapping of compute and communication, and the ability to exploit low-level optimization capabilities provided by the Primitive API. 

\spara{NCCL vs MSCCL.}
MSCCL is faster than NCCL in most cases, and all the benefit comes from better algorithms customized for our environments. The gain is especially big for small message sizes, where MSCCL uses an all-pairs algorithm (which is not supported by NCCL), while NCCL uses the ring algorithm that is worse in terms of latency. For very large message sizes $\ge$ 256MB in multi-node cases, MSCCL achieves higher bandwidth than NCCL because it uses a hierarchical algorithm (2PH), which is more bandwidth-efficient.

\spara{MSCCL vs \mscclpp{}.}
\mscclpp{} provides additional benefits over MSCCL by enabling more efficient variants of those algorithms.
For example, for single-node 1KB - 16KB messages in Figure~\ref{fig:all-reduce-a100}, both MSCCL and \mscclpp{} use the 1PA algorithm. For 1KB, we observe that \mscclpp{} cuts the latency by 47\% (from 9.5\usec{} to 5.0\usec{}), showing that the minimum overhead of the communication stack is substantially reduced. Larger messages from 32KB in single-node use various versions of the 2PA algorithm. Especially from 1MB to larger, efficient bandwidth utilization starts to matter, and \mscclpp{} shows better scalability than MSCCL. For 1GB in single-node, \mscclpp{} uses \texttt{PortChannel} that is not supported by NCCL/MSCCL within a single node. In this case, we observe that \texttt{PortChannel} achieves 6.2\% higher bandwidth than \texttt{MemoryChannel}. For 2-node and 4-node results in Figure~\ref{fig:all-reduce-a100}, the 2PH algorithm is used in all cases. \mscclpp{} shows substantial gains for both small and large messages.

\spara{DSL vs Primitive.}
From our 9-month experience using the \mscclpp{} DSL API, we find that it reduces development time from weeks to days, compared to using the Primitive API directly. The DSL helps automate initialization, check for mistakes, and provide performance analysis. We have implemented both the DSL and Primitive versions up to 2-node experiments to verify that their performance is similar, while 4-node experiments are done only using the DSL.
As the DSL introduces a runtime interpreter executor, 
DSL versions perform \textbf{3\%} worse than the Primitive versions on average, and are up to \textbf{18\%} worse
\blue{in one corner case. Despite performance overheads, \mscclpp{} DSL is useful for quick prototyping of collective communication with easy-to-understand algorithm description.}

\subsection{Cross-Hardware Support}\label{ssec:cross-hw}

\begin{figure}[t]
    \centering
    \includegraphics[width=\columnwidth]{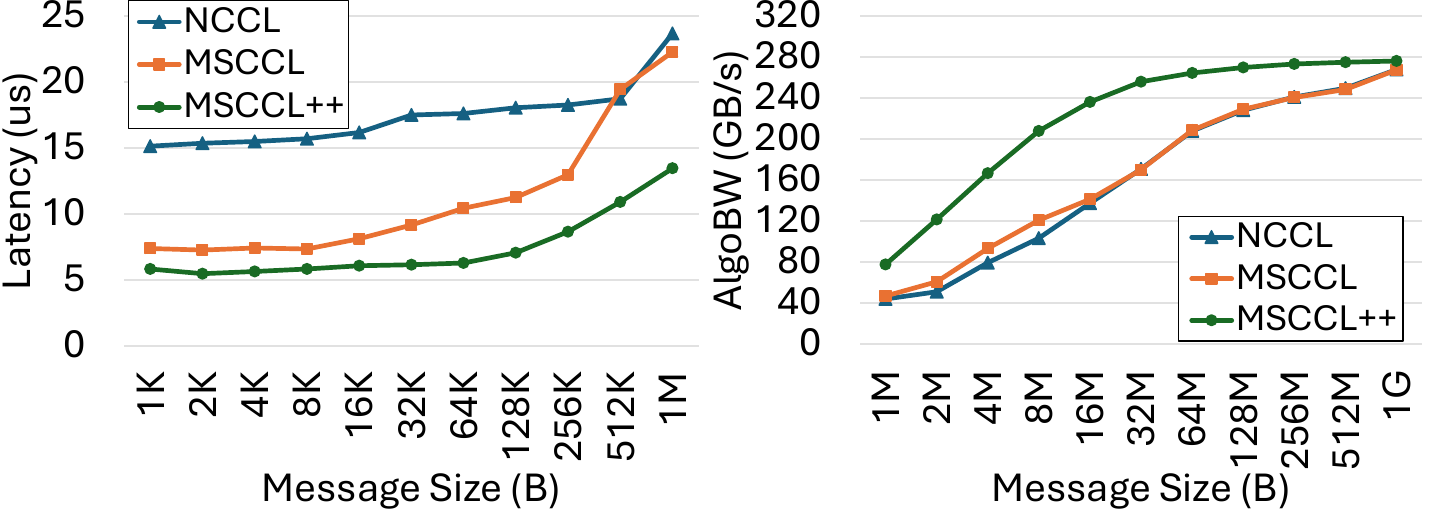}
    \caption{AllReduce, H100, single-node (8 GPUs).}
    \label{fig:all-reduce-h100-nvls}
\end{figure}

\begin{figure}[t]
    \centering
    \ifmscclpp
    \includegraphics[width=\columnwidth]{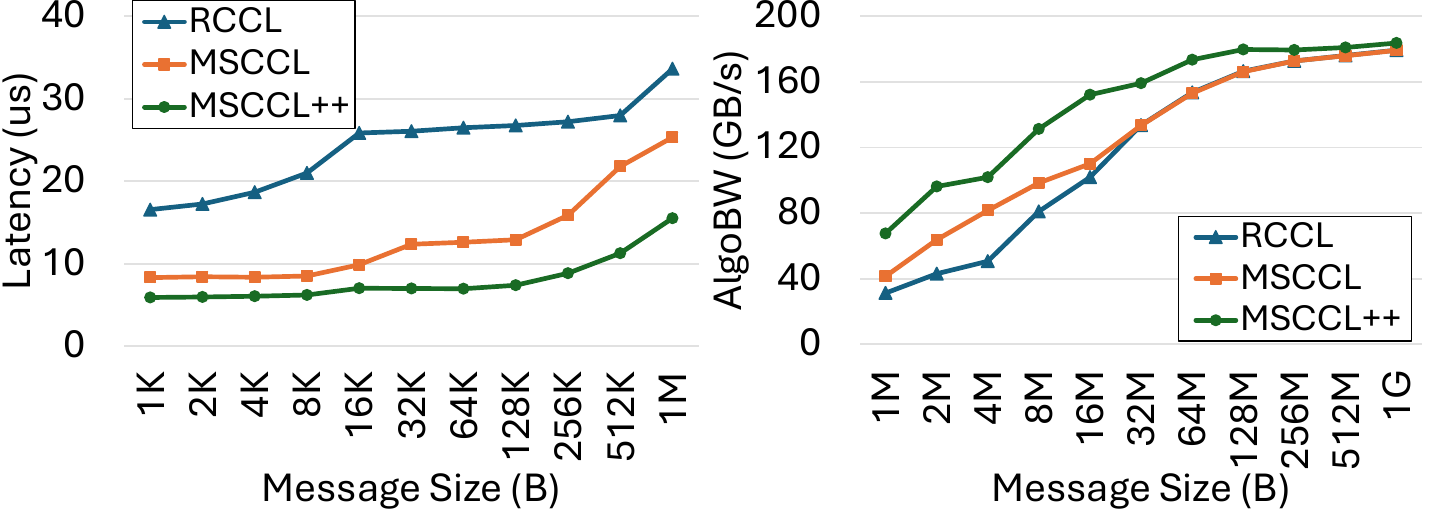}
    \else
    \includegraphics[width=\columnwidth]{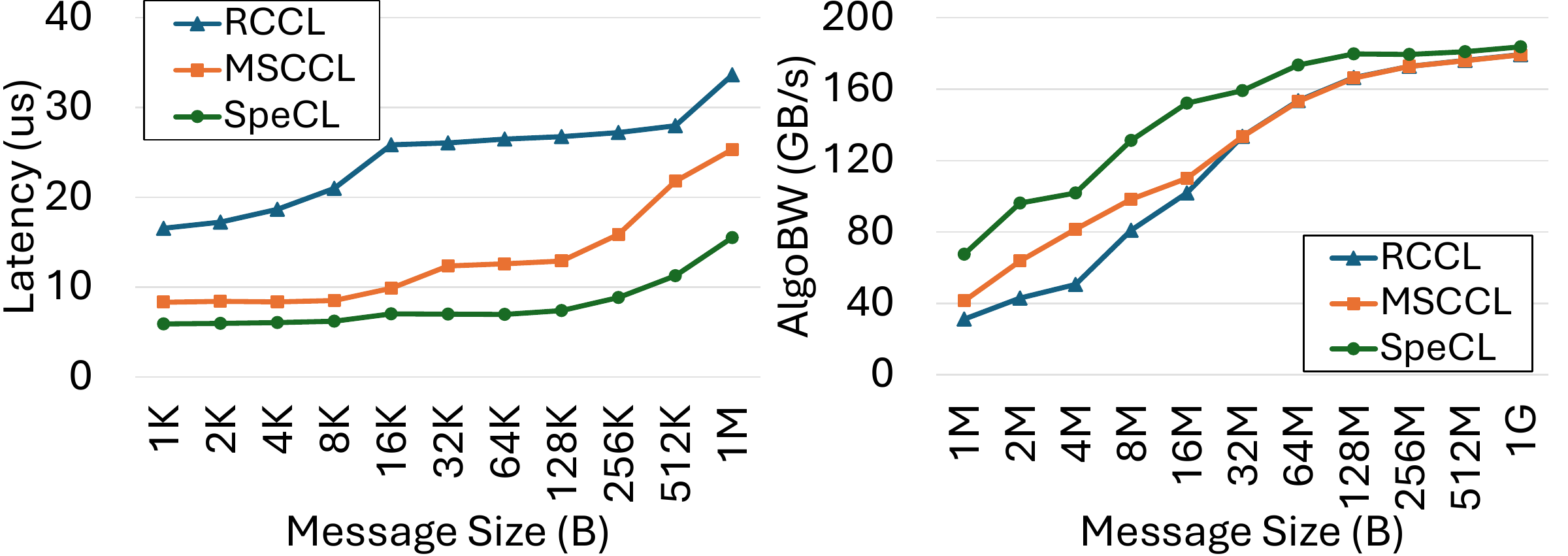}
    \fi
    \caption{AllReduce, MI300x, single-node (8 GPUs).}
    \label{fig:all-reduce-mi300x}
\end{figure}

\blue{To show the effectiveness of \mscclpp{} on different hardware platforms,} Figure~\ref{fig:all-reduce-h100-nvls} and \ref{fig:all-reduce-mi300x} compare the single-node AllReduce performance on H100 and MI300x.

\spara{H100.} \mscclpp{} outperforms NCCL and MSCCL by up to \textbf{2.8x} and \textbf{1.6x} for small messages, and by up to \textbf{2.4x} and \textbf{2.0x} for large messages, respectively. Our 2PA implementation made efficient use of H100's NVSwitch via the \texttt{SwitchChannel} interface, which delivers most of the benefit shown for large message sizes. Due to hardware acceleration, we observe up to 56$\%$ higher bandwidth by using \texttt{SwitchChannel} compared with an equivalent \texttt{MemoryChannel} implementation. 
In particular, our \texttt{SwitchChannel} code is only \textbf{15 lines of Python code} using our DSL, which simply calls \texttt{reduce} and \texttt{broadcast} element-wise in a loop. 
\blue{This shows that \mscclpp{} offers simple yet efficient interfaces, and a lot of unnecessary overhead in existing stacks can be avoided by using \mscclpp{}.}

\spara{MI300x.} \mscclpp{} outperforms RCCL and MSCCL by up to \textbf{3.8x} and \textbf{1.9x} for small messages, and by up to \textbf{2.2x} and \textbf{1.6x} for large messages, respectively. Our algorithm implementations take special considerations to best utilize the Infinity Fabric that peer-to-peer connects all GPUs in a node, unlike NVLink that connects all GPUs to a switch. This means, for best link utilization, we need to copy data to all peers at the same time as much as possible, unlike for NVIDIA GPUs where we can copy data to each peer one-by-one back-to-back. Using the \specl{} DSL, this is as easy as changing the order of two nested for loops, one that walks through data elements and the other that walks through peers.

\begin{figure}[t]
    \centering
    \includegraphics[width=\columnwidth]{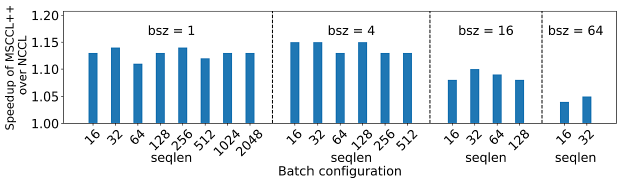}
    \caption{Speedup \blue{of \mscclpp{} over NCCL for Llama3-70b} decodes with tensor parallelism = 8.}
    \label{fig:e2e-decode-a100-1node}
\end{figure}

\subsection{LLM Inference}\label{ssec:llm-inference}

\noindent\textbf{LLM Inference with vLLM.} 
We evaluate the benefit of \specl{} in end-to-end distributed inference of a popular LLM, \blue{Llama3-70b~\cite{llama-3}}
using the LLM inference library, vLLM~\cite{kwon2023efficient} (v0.3.3). We modified vLLM to use \specl{} instead of NCCL for the AllReduce required in tensor parallelism and obtain the time taken for prefill and decode in an offline inference of a batch.
The model is distributed over all eight GPUs of a single-node A100-80G machine with tensor parallelism of 8 and CUDA graphs for decodes are enabled for performance.
Figure~\ref{fig:e2e-decode-a100-1node} shows that \specl{} is on average 1.11$\times$ faster than NCCL in decode latency for a range of batch configurations. Here $bsz$ denotes the number of batched requests, while $seqlen$ denotes the number of tokens in each request.
The reduction in decode time aligns perfectly with what we expect from our standalone AllReduce evaluation in Section~\ref{sssec:ccl-perf}.
Since the computation time for prefills is higher than decodes, the communication time improvements with \specl{} do not show up prominently, and we see similar or up to 1.06$\times$ faster prefill for different batch configurations.
\blue{A prior work~\cite{patel2024splitwise} has shown that in production traces, very few active tokens reside in a batch, and for most requests, the majority of end-to-end time is spent in the decode phase.
Thus, the performance improvements by \specl{} are expected to translate well to real-world workloads.}

vLLM also has a custom hand-written AllReduce kernel for a single node. For different message sizes, \mscclpp{}'s AllReduce performs similar or up to $3\times$ faster than the custom kernel, with a geomean improvement of $1.4\times$. For end-to-end decode latency, \mscclpp{} performs similar or up to $1.11\times$ (on average $1.04\times$) faster than inference with the custom AllReduce kernel.

\begin{figure}[t]
    \centering
    \ifmscclpp
    \includegraphics[width=\columnwidth]{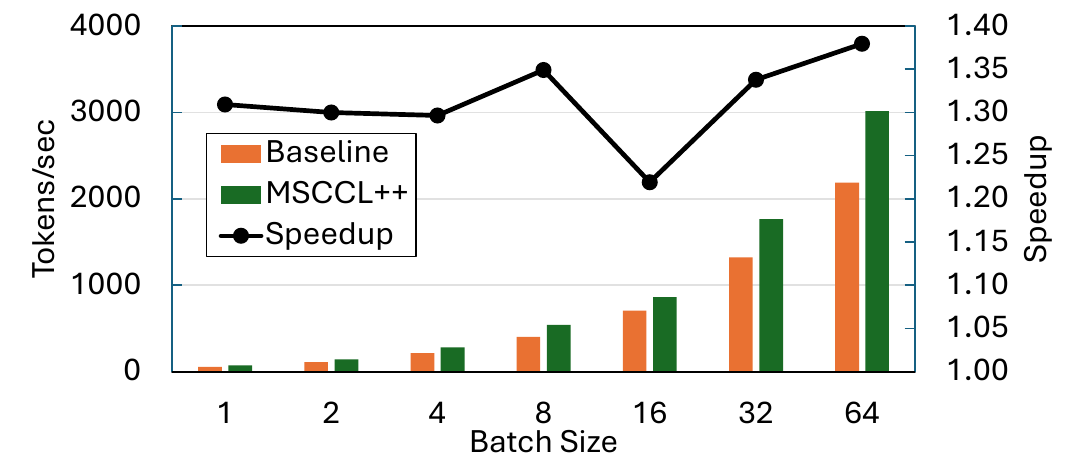}
    \else
    \includegraphics[width=\columnwidth]{figs/5-sglang-deepseek-decode.pdf}
    \fi
    \caption{DeepSeek-V3 decode throughput using SGLang on two H100 nodes (16 GPUs) with tensor parallelism (1024 input tokens and 1024 output tokens per batch).}
    \label{fig:sglang-mscclpp}
\end{figure}

\noindent\textbf{LLM Inference with SGLang.} 
SGLang~\cite{sglang} is a popular LLM inference framework that has adopted \mscclpp{} for accelerating AllReduce. 
We evaluated end-to-end distributed inference of a 
recent LLM, DeepSeek-V3~\cite{deepseek-v3}, using SGLang
on two H100 nodes with tensor parallelism of 16. 
SGLang has a custom AllReduce kernel but that 
is limited to a single node. 
\blue{Before adopting \mscclpp{}, 
SGLang was either using NCCL or a custom all-reduce kernel (depending on the platform and input size) by default, so that is our baseline.} 
Figure~\ref{fig:sglang-mscclpp} shows the decode throughput 
of the baseline and \mscclpp{}, and the speedup in 
decode throughput due to \mscclpp{}. 
\mscclpp{} is consistently faster and on average 1.31$\times$ faster than NCCL. 

\begin{figure}[t]
    \centering
    \ifmscclpp
    \includegraphics[width=\columnwidth]{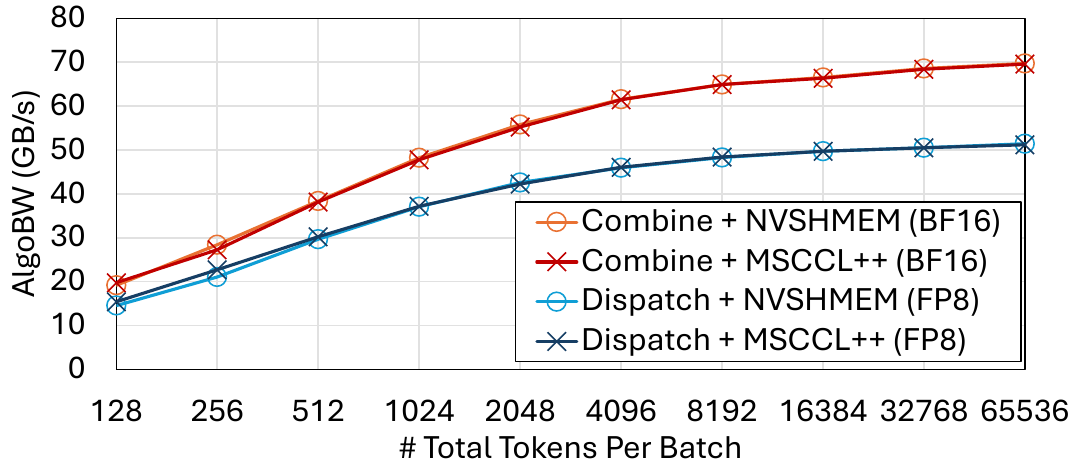}
    \else
    \includegraphics[width=\columnwidth]{figs/5trim-deepep-specl.pdf}
    \fi
    \caption{DeepEP on two H100 nodes (16 GPUs) with expert parallelism (DeepSeek-V3 settings: hidden size 7168, top-k 8, 256 total experts, FP8 dispatch and BF16 combine).}
    \label{fig:deepep-mscclpp}
\end{figure}

\noindent\textbf{Expert Parallelism with DeepEP.} DeepEP~\cite{deepep} is an implementation of expert parallelism for Mixture-of-Experts (MoE) layers, which is proposed by DeepSeek-V3~\cite{deepseek-v3} for both training and inference. The original code leverages a hardware feature called InfiniBand GPUDirect Async (IBGDA)~\cite{ibgda} (featured by NVSHMEM~\cite{nvshmem}) that implements an InfiniBand networking stack inside GPU, so that GPU can directly communicate with NIC bypassing CPU. However, this feature is only available on NVIDIA GPUs combined with Mellanox NICs at the time of writing. To make it portable to other environments, we integrate \mscclpp{} into DeepEP to replace the IBGDA-based networking functions into \texttt{PortChannel} API calls, which uses the RDMA networking stack running on CPU instead (see Figure~\ref{fig:mscclpp-ib}).\footnote{\blue{IBGDA can be supported by the \mscclpp{} \texttt{PortChannel} API in the future.}}
\blue{It took a \textbf{couple of weeks for one developer} to add this support, but most of that time was spent in understanding DeepEP code. We evaluated DeepEP using its official benchmark and setup. 
IBGDA was enabled for all experiments, which was leveraged by NVSHMEM, but not by \mscclpp{}.
Figure~\ref{fig:deepep-mscclpp} demonstrates that the two implementations exhibit no noticeable performance difference.
Unlike NCCL, \mscclpp{} is flexible enough to implement DeepEP-like applications efficiently.
The IBGDA stack implementation in DeepEP is less portable and hard to maintain, while \mscclpp{}’s \texttt{PortChannel} simplifies this code drastically while being portable and performant.}

\subsection{Portability and Implementation Effort}\label{sssec:quick-support}

\mscclpp{} is designed to be portable. Code using \mscclpp{} collectives, algorithms using the \mscclpp{} DSL, and kernels using the \mscclpp{} primitives are portable as long as the underlying \mscclpp{} Primitive API supports that hardware. 
Our design makes it easy to add support for new hardware. 
As an example, 
\mscclpp{} initially supported only NVIDIA GPUs; to add support for AMD MI300x GPUs, our implementation took only \blue{\textbf{7 weeks for one developer}}: 3 weeks for basic AMD GPU support and 4 weeks to develop new AllReduce algorithms that outperform RCCL/MSCCL for message sizes of 1KB - 1GB. 
\blue{We counted the lines of difference between before and after introducing support for the new MI300x architecture: RCCL 35480 vs \mscclpp{} 1307 (27$\times$ smaller).}
It is especially remarkable that the AMD-specific code in \mscclpp{} (excluding makefiles and algorithms) is fewer than 10 lines of code, while RCCL, the hard-forked NCCL for AMD GPUs, is substantially diverged from NCCL. \mscclpp{} could achieve this because the low-level API of AMD GPUs (i.e., HIP) is almost the same as that of NVIDIA GPUs (i.e., CUDA), and the \mscclpp{} Primitive API is only a shallow abstraction on top of the low-level API. This significantly reduces the effort to maintain and develop \mscclpp{}.

The \mscclpp{} design makes it easy to extend the Primitive API to support new hardware features.
For example, NVIDIA NVLink 4.0 introduced support for \texttt{multimem} instructions for switch-based aggregation and multicast. Since this feature introduced a very different concept from previous interfaces (i.e., \texttt{PortChannel} or \texttt{MemoryChannel}), 
we developed the \texttt{SwitchChannel} interface to support this feature. The development took only \blue{\textbf{8 weeks for two developers}}, including learning the basic usage of this feature, abstracting the feature as a new type of channel (i.e., \texttt{SwitchChannel}), and finally developing a new AllReduce algorithm using \texttt{SwitchChannel} that outperforms NCCL/MSCCL by more than $2.2\times$ on average for message sizes of 1KB - 1GB.

\section{Related Work}

\spara{GPU Communication Primitives:}
NVSHMEM~\cite{nvshmem} is an extension of the OpenSHMEM~\cite{openshmem} API for NVIDIA GPUs for providing a shared-memory abstraction. In contrast, \mscclpp{} provides performance-preserving portable abstractions and higher-level DSL for developing efficient communication collective algorithms. Concurrent to our work, NVSHMEM recently introduced support for inter-node communication (after \mscclpp{} was open sourced). \mscclpp{} provides different interconnect abstractions, namely \texttt{PortChannel}, \texttt{MemoryChannel}, and \texttt{SwitchChannel}, to enable optimized implementations. As an example, NVSHMEM’s \texttt{put} always uses thread-copy when it is feasible, so users cannot choose DMA-copy on NVLink. Similarly, NVSHMEM exposes raw \texttt{multimem} capabilities by exposing underlying memory pointers and users need to implement their kernels using PTX instructions. With an eye towards portability, \texttt{SwitchChannel} abstracts low-level details with a uniform synchronization mechanisms. 
Finally, NVSHMEM does not provide an API like MemoryChannel’s LL protocol. It may be possible to implement such an API using NVSHMEM, but that would add burden on users.

ARK~\cite{ark} is another work that proposes a GPU-side control plane for communication, but it is implemented as a monolithic end-to-end ML system rather than a standalone communication library.

\spara{DSL:}
Cowan et al. \cite{msccl} introduced MSCCLang, a DSL to specify collective communication algorithms specialized for a message size and network topology. 
MSCCLang is built on top of NCCL and RCCL send-recv abstraction, so is limited to two-sided, synchronous communication. 
Our \mscclpp{} DSL provides a one-sided, asynchronous API 
that enables expressing more efficient communication algorithms.
Due to this, \mscclpp{} is faster than MSCCL, which is built using MSCCLang DSL, as shown in our evaluation.

\spara{Communication Libraries:}
NVIDIA Collective Communications Library (NCCL)~\cite{nccl} and AMD ROCm Collective Communications Library~\cite{rccl} (RCCL) are vendor supplied libraries for NVIDIA and AMD GPUs respectively. 
We showed that \mscclpp{} Collectives perform better than both for AI inference workloads. 
RCCL has since adopted \mscclpp{}.

\spara{Communication Algorithms:}
SCCL~\cite{cai2021synthesizing}, TACCL~\cite{taccl}, and TE-CCL~\cite{te-ccl} aim to accelerate GPU collective communication by synthesizing efficient data transfer algorithms using the NCCL and RCCL send-recv abstraction. 
Their techniques are orthogonal and their performance may 
improve if they generate algorithms using the \mscclpp{} Primitives API.

\spara{LLM Inference Frameworks:}
LLM inference frameworks like SGLang~\cite{sglang}, DeepEP~\cite{deepep}, vLLM~\cite{kwon2023efficient}, and TensorRT-LLM~\cite{trt-llm} 
implement custom kernels for achieving high performance communication. However, the custom implementations are not general-purpose and often only limited to specific interconnects like NVLink or xGMI. \mscclpp{} not only performs similar to their custom implementations on their corresponding interconnects, but also supports efficient communication across multiple types of interconnects. 
SGLang has since adopted \mscclpp{}.

\section{\blue{Conclusion and Future Work}}
\specl{} is a novel GPU communication stack designed for high-performance AI applications.
By exposing the primitive communication functionalities as straightforward user interfaces, \mscclpp{}
enables fine-grained optimizations for GPU experts, while also providing higher-level interfaces for quick optimizations. Moreover, such a design can reduce the overall development and optimization effort for GPU communication, and accelerates adoption of fast evolving hardware technologies. By implementing collective communication using the proposed \specl{} interfaces, we can achieve up to $5.4\times$ speedup for standalone collectives, and up to $1.38\times$ speedup for end-to-end AI inference.
\blue{\mscclpp{} could be used for accelerating AI training workloads too, but we leave that for future work.}
\blue{\mscclpp{} could also evolve into a general I/O abstraction for AI accelerators in the future; e.g., by supporting non-GPU accelerators and file system I/O transports like BlitzScale~\cite{blitzscale}.}
\label{lastbody}

\bibliographystyle{ACM-Reference-Format}
\balance
\bibliography{reference}

\clearpage

\appendix
\section{Artifact Appendix}

\subsection{Abstract}


The artifact includes implementations of the three APIs described in this paper, namely the Primitive API (C++/CUDA) for low-level optimization of GPU communication, the DSL API (Python) that is implemented over the Primitive API and enables high-level description of custom collective communication algorithms, and the Collective API (C++) that implements the NCCL/RCCL API as-is over the Primitive API and the DSL API. The Collective API can be used to reproduce collective communication results in Section~\ref{sssec:ccl-perf} and \ref{ssec:cross-hw}.

\subsection{Artifact check-list (meta-information)}


{\small
\begin{itemize}
  \item {\bf Program: } The latest version of \texttt{nccl-tests} (or \texttt{rccl-tests}). Publicly available and not included.
  \item {\bf Run-time environment: } Linux; Ubuntu 20.04 or later is recommended. Docker is recommended, otherwise need manual installations of dependencies such as CUDA (or ROCm), Python3, MPI, and optionally OFED for multi-node experiments.
  \item {\bf Hardware: } 8 datacenter-grade GPUs. Optionally RDMA NICs for multi-node experiments.
  \item {\bf Execution: } Sole user recommended for best performance.
  \item {\bf Metrics: } Latency and bandwidth are reported.
  \item {\bf Output: } Console output with reported metrics.
  \item {\bf Experiments: } Step-by-step guide in documentation.
  \item {\bf How much disk space required (approximately)?: } <1GB for the source code and a build; 10GB for a Docker image with all dependencies.
  \item {\bf How much time is needed to prepare workflow (approximately)?: } 10 minutes
  \item {\bf How much time is needed to complete experiments (approximately)?: } 1 hour
  \item {\bf Publicly available?: } Yes
  \item {\bf Code licenses (if publicly available)?: } MIT License
  \item {\bf Archived (provide DOI)?: } 10.5281/zenodo.18334822
\end{itemize}
}

\subsection{Description}

\subsubsection{How to access}

The source code is publicly available on GitHub at \url{https://github.com/microsoft/mscclpp}. The repository is sized less than 1MB, and a build would be a few hundreds of MB.

\subsubsection{Hardware dependencies}

At least 2 GPUs are needed for functionality tests, and 8 or more are needed for reproducing numbers in Section~\ref{sssec:ccl-perf} and \ref{ssec:cross-hw}. Datacenter-grade GPUs with high-bandwidth interconnects (NVLink or Infinity Fabric) are recommended (see Table~\ref{tbl:env}). Optionally for multi-node experiments, RDMA NICs with InfiniBand or RoCE configurations are required.

\subsubsection{Software dependencies}

CUDA (or ROCm), Python3, and MPI are required for artifact evaluation. Optionally OFED is required for multi-node experiments. See details in the "Prerequisites" section in \texttt{docs/quickstart.md}.



\subsection{Installation}

\begin{enumerate}
    \item Download and run a provided Docker image. This step can be skipped if the host environment is preferred. See "Docker Images" section in \texttt{docs/quickstart.md}.
    
    \item Build from the source. See the step-by-step guide in "Install from Source" section in \texttt{docs/quickstart.md}. If the environment is not configured for multi-node execution, we recommend to disable InfiniBand/RoCE features via a CMake option \texttt{-DMSCCLPP\_USE\_IB=OFF}.
    
    \item Build and run \texttt{unit\_tests} and \texttt{mp\_unit\_tests} for functionality testing. See the step-by-step guide in "Unit Tests" section in \texttt{docs/quickstart.md}.
\end{enumerate}


\subsection{Evaluation and expected results}

For evaluation, install the latest version of \texttt{nccl-tests} (for NVIDIA GPUs) or \texttt{rccl-tests} (for AMD GPUs), which are publicly available at \url{https://github.com/NVIDIA/nccl-tests} and \url{https://github.com/ROCm/rccl-tests}, respectively. MPI should be enabled during their compilation (set environment variable \texttt{MPI=1} during \texttt{make}).

After that, follow the "NCCL/RCCL Benchmark over MSCCL++" section in \texttt{docs/quickstart.md} to run \texttt{nccl-tests} or \texttt{rccl-tests} using the \mscclpp{} Collective API. The testing binaries will print latency and bandwidth numbers on the console, similar to what presented in Figure~\ref{fig:all-reduce-a100}, \ref{fig:all-gather-a100}, \ref{fig:all-reduce-h100-nvls}, and \ref{fig:all-reduce-mi300x}, depending on the GPU type and the number of GPUs. Please refer to the \texttt{README.md} of \texttt{nccl-tests} for detailed usages.

\subsection{Experiment customization}

The implementation of \mscclpp{} Collective API can be customized using two different interfaces.
\begin{itemize}
  \item {\bf Using the Primitive API.} Users can write their own GPU communication kernel using the \mscclpp{} Primitive API and let the Collective API invoke the custom kernel. We provide a simple example in \\
  \texttt{docs/guide/customized-algorithm-with-nccl-api.md}.
  \item {\bf Using the DSL API.} Users can leverage our Python DSL to describe a collective communication algorithm that can be lowered and executed by the Collective API. We provide a simple example in \texttt{docs/dsl\_quick\_start.md}.
\end{itemize}

\subsection{Notes}

All documentation is also publicly available at \url{https://microsoft.github.io/mscclpp}.

\subsection{Methodology}

Submission, reviewing and badging methodology:

\begin{itemize}
  \item \url{https://www.acm.org/publications/policies/artifact-review-and-badging-current}
  \item \url{https://cTuning.org/ae}
\end{itemize}

\label{lastpage}
\clearpage

\end{document}